\documentclass[10pt]{article}

\usepackage{cite}
\usepackage{amsmath,amssymb,amsfonts}
\usepackage{graphicx}
\usepackage{textcomp}
\usepackage{xcolor}
\usepackage{booktabs}
\usepackage{url}
\usepackage{multirow}
\usepackage{array}
\usepackage{tikz}
\usepackage{pgfplots}
\pgfplotsset{compat=1.18}
\usetikzlibrary{arrows.meta, positioning, calc, patterns, decorations.pathreplacing}
\usepackage{tabularx}
\usepackage[margin=1in]{geometry}

\title{Deep Learning-Based Physical Layer Authentication\\
	Using 5G NR Sounding Reference Signals:\\
	A Temporal Generalization Study on Real Testbed Data}

	\author{
	Sachinkumar B. Mallikarjun$^{1}$, Marvin Reski$^{1}$, Andreas Weinand$^{1}$, and Hans D. Schotten$^{1,2}$ \\
	\\
	$^{1}$Division of Wireless Communications and Radio Navigation, \\
	Department of Electrical and Computer Engineering, \\
	RPTU University Kaiserslautern-Landau, Kaiserslautern, Germany \\
	\texttt{\{mallikar, marvin.reski, andreas.weinand, schotten\}@rptu.de}  \\
	$^{2}$German Research Center for Artificial Intelligence (DFKI), Kaiserslautern, Germany \\
	\texttt{schotten\}@dfki.de}
}

\begin{document}
	\maketitle
	
		\begin{center}
		\textit{Submitted to the European Wireless 2026 conference}
	\end{center}
	% ══════════════════════════════════════════════════════════════════════
	%  ABSTRACT
	% ══════════════════════════════════════════════════════════════════════
	\begin{abstract}
		Physical Layer Authentication (PLA) exploits the spatial uniqueness of wireless channel characteristics in order to authenticate devices without recourse to higher-layer cryptographic protocols, which remain vulnerable to key compromise. This paper reports a comprehensive PLA system constructed on 5G New Radio (NR) Sounding Reference Signals (SRS) extracted from a real OpenAirInterface (OAI) testbed operating in band~n78 (3.5\,GHz) with 40\,MHz bandwidth and 30\,kHz subcarrier spacing. The proposed approach extracts a 2{,}531-dimensional feature vector per SRS probe, combining per-subcarrier channel state information (1{,}248 amplitude and 1{,}247 differential-phase coefficients), power delay profile taps, delay spread, Doppler statistics, and nonlinear dynamics indicators. A deep one-dimensional Residual Network (1D-ResNet) augmented with Squeeze-and-Excitation (SE) attention blocks is employed to classify each probe as either legitimate or spoofed. Evaluation is conducted on 20{,}317 over-the-air SRS probes acquired across four measurement sessions using a USRP~B210 software-defined radio as the legitimate device and a commercial mobile handset as the attacker. Under a strict \emph{chronological} train/validation/test split that eliminates temporal leakage, an Equal Error Rate (EER) of \textbf{3.92\%} is attained, with AUC\,=\,0.962 on the held-out test set, and an authentication latency of less than 0.1\,ms per probe, which is compatible with 5G Ultra-Reliable Low-Latency Communications (URLLC) requirements.% It is further shown that a na\"ive random split inflates the reported performance to an EER of 2.58\%, thereby highlighting the importance of temporally disciplined evaluation in PLA research.
	\end{abstract}

	% ══════════════════════════════════════════════════════════════════════
	%  I. INTRODUCTION
	% ══════════════════════════════════════════════════════════════════════
	\section{Introduction}
	\label{sec:intro}
	
	Physical Layer Authentication (PLA) has emerged as a promising complement to cryptographic device authentication in wireless networks~\cite{wu2015pla,xie2019channel,xiao2008physical}. In contrast to higher-layer protocols, which rely on shared secrets that may become vulnerable in the presence of advances in quantum computing, PLA exploits the inherent spatial uniqueness of wireless propagation channels. Each transmitter--receiver pair experiences a distinct channel impulse response determined by the surrounding physical multipath environment, which permits the receiver to \emph{fingerprint} legitimate devices on the basis of their observed channel characteristics.
	
	Fifth-generation New Radio (5G~NR) introduces several features that substantially enhance PLA capability. The Sounding Reference Signal (SRS), standardised in 3GPP~TS~38.211~\cite{3gpp38211}, provides dedicated uplink channel estimation pilots with configurable bandwidth, periodicity, and comb structure. Under wideband operation at 40\,MHz in band~n78, a single SRS probe spans 1{,}248 active subcarriers, each of which carries an independent complex channel coefficient. This yields substantially richer fingerprinting material than the narrowband reference signals employed in LTE.
	
	Despite these advantages, the majority of existing PLA studies rely on either simulated channel models or narrowband LTE and WiFi measurements~\cite{liu2017pls,xiao2018iot,wang2019pls}. A further methodological concern is that many recent deep learning-based PLA evaluations employ random train/test splits, which introduce temporal leakage: consecutive SRS probes drawn from the same session share highly correlated channel realisations, and the random distribution of such probes across the training and test sets artificially inflates the reported accuracy~\cite{hoang2024pla}. This methodological issue has received limited attention in the literature.
	
	The contributions of this paper are summarised as follows.
	\begin{enumerate}
		\item A complete PLA pipeline is designed that extracts a 2{,}531-dimensional multi-domain feature vector from real 5G~NR SRS channel estimates, encompassing per-subcarrier amplitude and differential phase, power delay profile, Doppler statistics, and nonlinear dynamics indicators.
		\item A deep 1D-ResNet classifier augmented with Squeeze-and-Excitation (SE) attention blocks is implemented and trained with mixup augmentation, label smoothing, and a cosine-annealing learning rate schedule.
		\item Validation is performed on 20{,}317 over-the-air SRS probes acquired from four measurement sessions on an OAI 5G~NR gNB (band~n78, 40\,MHz bandwidth, 30\,kHz subcarrier spacing), with two physically distinct transmit devices: a USRP~B210 SDR and a commercial mobile handset.
		\item A rigorous comparison between a chronological train/validation/test split and random split is reported, which quantifies the performance inflation induced by temporal leakage.
	\end{enumerate}
	
	The remainder of this paper is organised as follows. Section~\ref{sec:related} reviews related work. Section~\ref{sec:system} describes the testbed and the data collection procedure. Section~\ref{sec:features} details the feature extraction pipeline. Section~\ref{sec:models} presents the authentication models. Section~\ref{sec:results} reports the experimental results, and Section~\ref{sec:conclusion} concludes the paper.
	
	% ══════════════════════════════════════════════════════════════════════
	%  II. RELATED WORK
	% ══════════════════════════════════════════════════════════════════════
	\section{Related Work}
	\label{sec:related}
	
	Channel-based PLA was introduced by Xiao et al.~\cite{xiao2008physical}, who proposed hypothesis testing on temporal channel variations in WiFi systems and reported authentication error rates in the range of 5--8\% under time-varying indoor channel conditions. Subsequent work by Liu et al.~\cite{liu2017pls} applied machine learning to LTE Channel State Information (CSI) for PLA in simulation, with reported EER values of approximately 5.2\%.
	
	The transition to deep learning yielded substantial improvements in authentication accuracy. Xiao et al.~\cite{xiao2018iot} employed a convolutional neural network (CNN) applied to WiFi CSI for Internet of Things (IoT) device authentication, attaining 3.8\% EER on real indoor traces acquired from commodity hardware. Wang et al.~\cite{wang2019pls} subsequently investigated 5G~NR PLA in a simulated environment using a Long Short-Term Memory (LSTM)-based detector and reported 2.1\% EER; however, the use of synthetic channel models limits the generalisability of such results to real deployments.
	
	More recently, Zha et al.~\cite{zha2025srs} proposed a cross-domain radio-frequency (RF) fingerprinting system that leverages 5G~NR SRS signals with a convolutional architecture, with a stated objective of robustness against channel variations. Their approach reports identification accuracy in excess of 95\% on a commercial 5G testbed; however, neither EER nor performance under strict temporal splits was reported. Lin et al.~\cite{lin2024rffi} surveyed deep learning methods for 5G RF fingerprint identification and remarked upon the persistent evaluation gap between simulation-based and real-data studies.
	
	The present work is distinguished from prior art in three respects: (i)~a full 5G~NR OAI gNB is employed with direct \texttt{T\_tracer} access to the raw uplink SRS channel estimates; (ii)~a 2{,}531-dimensional feature vector is extracted across six signal domains; and (iii)~evaluation is explicitly conducted under a chronological split in order to provide temporally honest performance metrics.
	
	% ══════════════════════════════════════════════════════════════════════
	%  III. SYSTEM MODEL AND TESTBED
	% ══════════════════════════════════════════════════════════════════════
	\section{System Model and Testbed}
	\label{sec:system}
	
	\subsection{5G NR Testbed Configuration}
	
	The experimental testbed comprises an OAI gNB and nrUE, each deployed on USRP~B210 software-defined radios. The gNB executes the full 5G~NR Layer~1 and Layer~2 stack and is configured for Time Division Duplex (TDD) operation in band~n78 (3.5\,GHz) with a channel bandwidth of 40\,MHz, subcarrier spacing of 30\,kHz, and a DDDSU frame pattern. The principal radio parameters are summarised in Table~\ref{tab:testbed}.
	
	\begin{table}[t]
		\centering
		\caption{5G NR Testbed Parameters}
		\label{tab:testbed}
		\begin{tabular}{@{}ll@{}}
			\toprule
			\textbf{Parameter} & \textbf{Value} \\
			\midrule
			Band                 & n78 (3.5\,GHz TDD) \\
			Channel bandwidth    & 40\,MHz \\
			Subcarrier spacing   & 30\,kHz \\
			Number of PRBs       & 106 \\
			FFT size             & 1536 \\
			Active subcarriers   & 1{,}248 (nonzero) \\
			SRS periodicity      & 80\,ms (12.5\,probes/s) \\
			TDD pattern          & DDDSU \\
			Antenna config.      & 1$\times$1 (SISO) \\
			gNB / UE platform    & USRP B210 + OAI 5G NR \\
			\bottomrule
		\end{tabular}
	\end{table}
	
	\subsection{T\_tracer CSI Extraction}
	
	OAI provides a real-time tracing facility (\texttt{T\_tracer}) which logs internal PHY-layer events as binary records. Each SRS reception at the gNB triggers the generation of four events:
	\begin{enumerate}
		\item \texttt{UL\_FREQ\_CHANNEL\_ESTIMATE}: the frequency-domain channel estimate $\hat{H}(f_k)$, stored as 1{,}536 packed \texttt{c16\_t} complex samples (1{,}248 of which are nonzero).
		\item \texttt{UL\_TIME\_CHANNEL\_ESTIMATE}: the time-domain channel impulse response $\hat{h}(\tau_l)$, stored as 3{,}072 \texttt{c16\_t} samples.
		\item \texttt{UL\_SNR\_ESTIMATE}: per-resource-block signal-to-noise ratio (SNR) estimates.
		\item \texttt{UL\_SRS\_TOA\_NS}: Time-of-Arrival (ToA), reported in nanoseconds.
	\end{enumerate}
	A Python parser was developed to decode these events from the binary format, including the handling of the \texttt{c16\_t} packed integer representation in which each 32-bit word encodes the in-phase component $I$ (lower 16~bits) and the quadrature component $Q$ (upper 16~bits) as Q15-scaled fixed-point values.
	
	\subsection{Data Collection}
	\label{sec:data}
	
	SRS traces were acquired across four measurement sessions spanning multiple days, with temporal separation between sessions deliberately introduced in order to capture channel variations arising from environmental changes (Table~\ref{tab:traces}). The legitimate device~(UE1) is a USRP~B210 SDR executing the OAI nrUE stack. The attacker~(UE2) is a commercial 5G mobile handset operated at varying positions. The use of entirely different hardware platforms ensures that the cross-device evaluation captures both channel and hardware fingerprint differences.
	
	During anchor sessions~2 and~3, the legitimate UE underwent multiple Radio Resource Control (RRC) reconnections, each of which was assigned a new temporary Radio Network Temporary Identifier (RNTI) by the gNB. Through manual inspection, it was verified that all RNTIs across these sessions belonged to the same physical UE1 device; consequently, no RNTI filtering was applied, and all SRS probes were retained as legitimate training data.
	
	\begin{table}[t]
		\centering
		\caption{Measurement Traces (SRS Probes)}
		\label{tab:traces}
		\footnotesize
		\begin{tabular}{@{}llrccc@{}}
			\toprule
			\textbf{Trace} & \textbf{Device} & \textbf{SRS}
			& \textbf{RNTIs} & \textbf{Dist.} & \textbf{Traffic} \\
			\midrule
			Trace\,1  & UE1 (USRP)  &  3{,}211 & 2  & 0.75\,m & Yes \\
			Anchor\,1 & UE1 (USRP)  &  2{,}358 & 1  & 1.0\,m  & No \\
			Anchor\,2 & UE1 (USRP)  &  7{,}429 & 35 & 1.0\,m  & Yes \\
			Anchor\,3 & UE1 (USRP)  &  6{,}808 & 9  & 1.0\,m  & Mixed \\
			\midrule
			Attack    & UE2 (Phone)  &    511   & ---  & varied & Yes \\
			\midrule
			\multicolumn{2}{@{}l}{\textbf{Total}} & \textbf{20{,}317} & & & \\
			\bottomrule
		\end{tabular}
	\end{table}
	
	% ══════════════════════════════════════════════════════════════════════
	%  IV. FEATURE EXTRACTION
	% ══════════════════════════════════════════════════════════════════════
	\section{Feature Extraction}
	\label{sec:features}
	
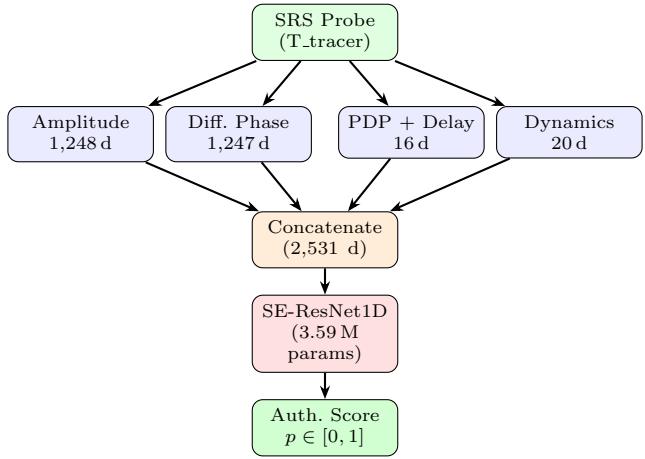
\begin{figure}[t]
		\centering
		\begin{tikzpicture}[
			block/.style={draw, rounded corners, minimum width=1.8cm,
				minimum height=0.65cm, font=\scriptsize, text width=1.7cm,
				align=center, fill=blue!8},
			arrow/.style={-{Stealth[length=1.8mm]}, thick},
			node distance=0.3cm and 0.3cm,
			]
			\node[block, fill=green!12] (srs) {SRS Probe\\(T\_tracer)};
			\node[block, below left=0.6cm and 1.3cm of srs]  (amp)
			{Amplitude\\1{,}248\,d};
			\node[block, below left=0.6cm and -0.8cm of srs]  (phase)
			{Diff.\ Phase\\1{,}247\,d};
			\node[block, below right=0.6cm and -0.8cm of srs] (pdp)
			{PDP + Delay\\16\,d};
			\node[block, below right=0.6cm and 1.3cm of srs]  (dyn)
			{Dynamics\\20\,d};
			\node[block, fill=orange!15, below=2.0cm of srs] (cat)
			{Concatenate\\$(2{,}531\;\text{d})$};
			\node[block, fill=red!12, below=0.35cm of cat] (model)
			{SE-ResNet1D\\(3.59\,M params)};
			\node[block, fill=green!18, below=0.35cm of model] (out)
			{Auth.\ Score\\$p\!\in\![0,1]$};
			\draw[arrow] (srs) -- (amp);
			\draw[arrow] (srs) -- (phase);
			\draw[arrow] (srs) -- (pdp);
			\draw[arrow] (srs) -- (dyn);
			\draw[arrow] (amp)   -- (cat);
			\draw[arrow] (phase) -- (cat);
			\draw[arrow] (pdp)   -- (cat);
			\draw[arrow] (dyn)   -- (cat);
			\draw[arrow] (cat)   -- (model);
			\draw[arrow] (model) -- (out);
		\end{tikzpicture}
		\caption{PLA feature extraction and classification pipeline. Six feature groups are concatenated into a 2{,}531-dimensional vector and supplied to a 1D-ResNet classifier augmented with SE attention.}
		\label{fig:pipeline}
	\end{figure}

	Six feature groups are extracted from each SRS probe, which together yield a 2{,}531-dimensional feature vector. The processing pipeline is illustrated in Fig.~\ref{fig:pipeline}.
	
	% ── Pipeline figure ───────────────────────────────────────────────────
	
	\subsection{Per-Subcarrier Amplitude (1{,}248 dimensions)}
	
	The magnitude at each subcarrier of the frequency-domain channel estimate $\hat{H}(f_k)$ is extracted across the $K\!=\!1{,}248$ active subcarriers,
	\begin{equation}
		a_k = |\hat{H}(f_k)|, \quad k = 0, 1, \ldots, K{-}1.
		\label{eq:amp}
	\end{equation}
	In contrast to coarser binning approaches that average over frequency groups, the retention of per-subcarrier resolution preserves the fine-grained frequency-selective patterns that are characteristic of each transmitter's analog front-end impairments and of its particular multipath channel.
	
	\subsection{Differential Phase (1{,}247 dimensions)}
	
	The raw carrier phase is dominated by carrier frequency offset (CFO) and timing offset, both of which are device-state-dependent rather than channel-dependent. For this reason, the \emph{differential phase} between adjacent subcarriers is extracted,
	\begin{equation}
		\Delta\phi_k = \angle\hat{H}(f_{k+1}) - \angle\hat{H}(f_k),
		\quad k = 0, \ldots, K{-}2.
		\label{eq:dphase}
	\end{equation}
	This operation removes the common linear phase slope introduced by timing offset, while preserving the multipath-induced phase structure that has been shown to carry discriminative device-level information~\cite{xie2019channel}.
	
	\subsection{Power Delay Profile and Delay Spread (16 dimensions)}
	
	The following features are derived from the time-domain channel impulse response $\hat{h}(\tau_l)$:
	\begin{itemize}
		\item Top-5 PDP tap powers and their delay indices (10\,d): the five strongest multipath components and their relative delays, which together characterise the propagation geometry.
		\item RMS delay spread $\tau_{\text{rms}}$ (1\,d):
		\begin{equation}
			\tau_{\mathrm{rms}} = \sqrt{
				\frac{\sum_l |\hat{h}(\tau_l)|^2\,(\tau_l - \bar{\tau})^2}
				{\sum_l |\hat{h}(\tau_l)|^2}
			}.
			\label{eq:delay}
		\end{equation}
		\item ToA (1\,d): as reported by the gNB timing-advance estimator in nanoseconds.
		\item Coherence bandwidth (1\,d): estimated as the reciprocal of $\tau_{\mathrm{rms}}$.
		\item Amplitude statistics (3\,d): the mean, standard deviation, and kurtosis of the subcarrier amplitudes.
	\end{itemize}
	
	\subsection{Doppler and Temporal Dynamics (8 dimensions)}
	
	The following features are computed over a sliding window of $W\!=\!20$ consecutive probes: the mean Doppler shift estimated from the decay rate of the channel autocovariance, the maximum and standard deviation of the Doppler estimates, and five additional temporal statistics, namely the coherence time, the amplitude entropy, and inter-probe correlation decay statistics.
	
	\subsection{Nonlinear Dynamics Indicators (12 dimensions)}
	
	In order to capture hardware-specific nonlinearities, which are inherently difficult to spoof, the following quantities are computed: the wavelet variance at eight dyadic scales, the sample entropy, the fractal dimension, the largest Lyapunov exponent, and the recurrence rate of the subcarrier amplitude time series. These features target analog impairments such as in-phase/quadrature (IQ) imbalance and power amplifier nonlinearity, which together form a device-specific RF fingerprint.
	
	\smallskip
	\noindent\emph{In total, $1{,}248 + 1{,}247 + 16 + 8 + 12 = 2{,}531$ features are extracted per SRS probe.}
	
	% ══════════════════════════════════════════════════════════════════════
	%  V. AUTHENTICATION MODELS
	% ══════════════════════════════════════════════════════════════════════
	\section{Authentication Models}
	\label{sec:models}
	
	\subsection{Deep Learning: SE-ResNet1D}
	
	A deep one-dimensional Residual Network~\cite{he2016resnet}, augmented with Squeeze-and-Excitation attention blocks~\cite{hu2018squeeze}, is employed. The architecture is defined as follows.
	
	\begin{itemize}
		\item \textbf{Stem:} Linear$(2531, 256)$ $\to$ BatchNorm $\to$ ReLU $\to$ Conv1d$(256, 256, k\!=\!7, s\!=\!2)$ $\to$ BatchNorm $\to$ ReLU $\to$ MaxPool$(k\!=\!3, s\!=\!2)$.
		\item \textbf{Body:} $6 \times$ SE-ResBlock1D, in which each block comprises two Conv1d$(256, 256, k\!=\!3)$ layers with batch normalisation, ReLU activation, and dropout ($p\!=\!0.2$), together with an SE module (reduction ratio of 16) that learns per-channel importance weights via global average pooling followed by a two-layer bottleneck.
		\item \textbf{Head:} AdaptiveAvgPool1d$(1)$ $\to$ Linear$(256, 256)$ $\to$ ReLU $\to$ Linear$(256, 2)$.
	\end{itemize}
	
	The total number of parameters is 3.59\,M, and the output is the softmax probability $p_{\text{legit}} \in [0,1]$.
	
	\noindent\textbf{Training protocol:}
	\begin{itemize}
		\item The Adam optimiser is employed with learning rate $10^{-3}$ and weight decay $10^{-4}$.
		\item A cosine-annealing learning rate schedule is applied over 100 training epochs.
		\item Label smoothing ($\varepsilon\!=\!0.1$) is employed in order to improve output calibration.
		\item Mixup augmentation ($\alpha\!=\!0.2$) is applied for regularisation.
		\item Gradient clipping is imposed at a maximum norm of $1.0$.
		\item Early stopping is triggered on the validation accuracy, with a patience of 15 epochs.
	\end{itemize}
	
	\subsection{Baseline: Pearson Correlation Threshold}
	
	As a classical baseline, the gNB computes a reference amplitude profile $\bar{\mathbf{a}}$ by averaging over $N_{\text{enroll}}$ calibration probes. Each test probe is authenticated if its Pearson correlation with the reference exceeds a threshold $\alpha$,
	\begin{equation}
		\rho(\mathbf{a}, \bar{\mathbf{a}}) =
		\frac{\sum_k (a_k - \bar{a})(\bar{a}_k - \overline{\bar{a}})}
		{\sqrt{\sum_k (a_k - \bar{a})^2}\;\sqrt{\sum_k (\bar{a}_k -
				\overline{\bar{a}})^2}}
		\;\geq\; \alpha,
		\label{eq:pearson}
	\end{equation}
	where the default threshold is set to $\alpha\!=\!0.85$.
	
	% ══════════════════════════════════════════════════════════════════════
	%  VI. EXPERIMENTAL EVALUATION
	% ══════════════════════════════════════════════════════════════════════
	\section{Experimental Results}
	\label{sec:results}
	
	\subsection{Evaluation Protocol}
	\label{sec:protocol}
	
	In order to prevent temporal leakage, which may arise when correlated consecutive probes from the same session appear in both the training and the test sets, a \textbf{chronological split} is adopted and applied independently within each trace:
	\begin{itemize}
		\item \textbf{Training (70\%):} the first 70\% of probes in temporal order.
		\item \textbf{Validation (15\%):} the subsequent 15\% of probes.
		\item \textbf{Test (15\%):} the final 15\% of probes.
	\end{itemize}
	This arrangement ensures that the model is evaluated on the \emph{temporally latest} data from each session, which was never observed during training or hyperparameter selection.
	
	The resulting split sizes are reported in Table~\ref{tab:splits}. A substantial class imbalance, with a legitimate-to-attack ratio of approximately 38:1, is noted; this reflects a realistic deployment scenario in which the majority of traffic is legitimate.
	
	\begin{table}[t]
		\centering
		\caption{Dataset Splits (Chronological)}
		\label{tab:splits}
		\begin{tabular}{@{}lrrr@{}}
			\toprule
			& \textbf{Train} & \textbf{Val} & \textbf{Test} \\
			\midrule
			Legitimate (UE1)  & 13{,}862 & 2{,}971 & 2{,}973 \\
			Attack (UE2)      &      357 &      77 &      77 \\
			\midrule
			\textbf{Total}    & 14{,}219 & 3{,}048 & 3{,}050 \\
			\bottomrule
		\end{tabular}
	\end{table}
	
	For the purposes of comparison, an identical model is additionally trained under a \textbf{random 70/15/15 split} (shuffled across all traces) in order to quantify the extent to which temporal leakage inflates the reported metrics.
	
	\subsection{Cross-Device Authentication Results}
	\label{sec:crossdevice}
	
	The primary authentication results obtained on the held-out test set are reported in Table~\ref{tab:results}.
	
	\begin{table}[t]
		\centering
		\caption{Authentication Performance on Held-Out Test Set}
		\label{tab:results}
		\begin{tabular}{@{}lcccc@{}}
			\toprule
			\textbf{Split Method} & \textbf{Accuracy} & \textbf{EER}
			& \textbf{AUC} & \textbf{Epochs} \\
			\midrule
			Chronological         & 97.48\% & 3.92\% & 0.962 & 16 \\
			Random (baseline)     & 99.74\% & 2.58\% & 0.991 & 32 \\
			\bottomrule
		\end{tabular}
	\end{table}
	
	Under the chronological split, the SE-ResNet1D classifier attains an EER of \textbf{3.92\%}, with a False Acceptance Rate (FAR) of 3.90\% and a False Rejection Rate (FRR) of 3.94\% at the EER operating point (threshold $\tau\!=\!0.947$). The random split, by contrast, yields a substantially lower EER of 2.58\%; this corresponds to a relative improvement of 34\% and is attributed to temporal leakage, since the distribution of consecutive probes from the same session across the training and test sets allows the model to exploit short-term channel correlation rather than to learn device-invariant fingerprints.
	
	The 1.34 percentage-point gap between the two splits quantifies the optimistic bias introduced by random splitting. It is accordingly recommended that future PLA work adopt chronological or session-level splits in order to avoid overestimating the generalisation performance of the authentication system.
	
	% ── DET curve figure ──────────────────────────────────────────────────
	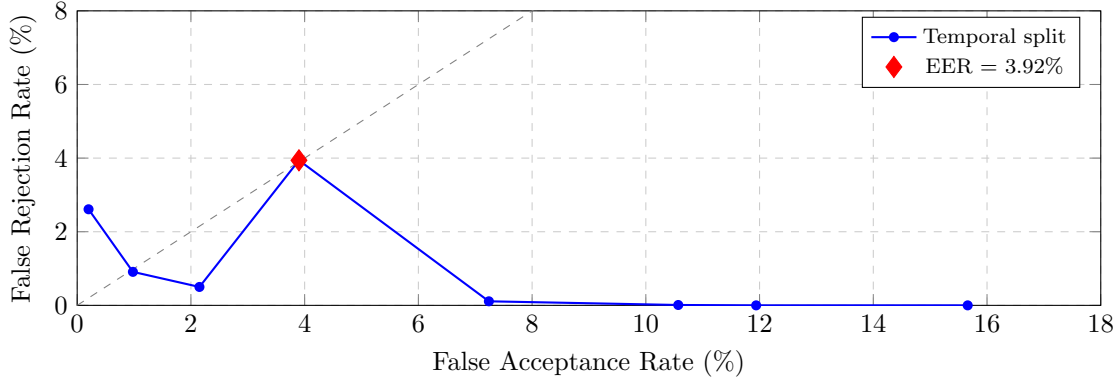
\begin{figure}[t]
		\centering
		\begin{tikzpicture}
			\begin{axis}[
				width=0.92\columnwidth,
				height=5.5cm,
				xlabel={False Acceptance Rate (\%)},
				ylabel={False Rejection Rate (\%)},
				xmin=0, xmax=18,
				ymin=0, ymax=8,
				grid=major,
				grid style={dashed, gray!40},
				mark options={solid},
				legend style={at={(0.98,0.98)}, anchor=north east,
					font=\footnotesize},
				]
				% DET curve data points (temporal split)
				\addplot[blue, thick, mark=*, mark size=1.5pt]
				coordinates {
					(0.20, 2.61)
					(0.98, 0.91)
					(2.15, 0.50)
					(3.90, 3.94)
					(7.24, 0.11)
					(10.57, 0.01)
					(11.94, 0.00)
					(15.66, 0.00)
				};
				\addlegendentry{Temporal split}
				
				% EER point
				\addplot[red, only marks, mark=diamond*, mark size=4pt]
				coordinates {(3.90, 3.94)};
				\addlegendentry{EER = 3.92\%}
				
				% Diagonal
				\addplot[gray, thin, dashed, domain=0:18]{x};
			\end{axis}
		\end{tikzpicture}
		\caption{Detection Error Tradeoff (DET) curve on the held-out test set (chronological split). The diamond marks the Equal Error Rate operating point at 3.92\%.}
		\label{fig:det}
	\end{figure}
	
	\subsection{Training Dynamics}
	
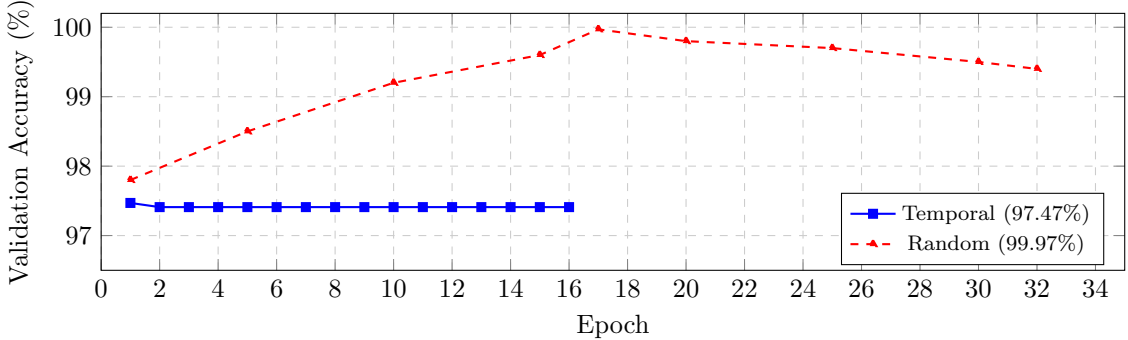
\begin{figure}[t]
		\centering
		\begin{tikzpicture}
			\begin{axis}[
				width=0.92\columnwidth,
				height=5.0cm,
				xlabel={Epoch},
				ylabel={Validation Accuracy (\%)},
				xmin=0, xmax=35,
				ymin=96.5, ymax=100.2,
				grid=major,
				grid style={dashed, gray!40},
				legend style={at={(0.98,0.02)}, anchor=south east,
					font=\footnotesize},
				]
				% Temporal split val acc
				\addplot[blue, thick, mark=square*, mark size=1.5pt]
				coordinates {
					(1,97.47) (2,97.41) (3,97.41) (4,97.41) (5,97.41)
					(6,97.41) (7,97.41) (8,97.41) (9,97.41) (10,97.41)
					(11,97.41) (12,97.41) (13,97.41) (14,97.41) (15,97.41)
					(16,97.41)
				};
				\addlegendentry{Temporal (97.47\%)}
				
				% Random split val acc (representative)
				\addplot[red, thick, dashed, mark=triangle*, mark size=1.5pt]
				coordinates {
					(1,97.8) (5,98.5) (10,99.2) (15,99.6) (17,99.97)
					(20,99.8) (25,99.7) (30,99.5) (32,99.4)
				};
				\addlegendentry{Random (99.97\%)}
			\end{axis}
		\end{tikzpicture}
		\caption{Validation accuracy during training. The temporal split plateaus early, whereas the random split attains near-perfect accuracy as a consequence of temporal leakage between adjacent probes.}
		\label{fig:training}
	\end{figure}

	The training and validation accuracy curves are shown in Fig.~\ref{fig:training}. Under the chronological split, the model converges rapidly, attaining its best validation accuracy at epoch~1 and terminating via early stopping at epoch~16. This behaviour is indicative of a temporal distribution shift between the training and validation data, which bounds the achievable accuracy. Under the random split, by contrast, training proceeds for 32~epochs and reaches 99.97\% validation accuracy, as temporally adjacent probes leak across the split boundary.
	
	% ── Training curves figure ────────────────────────────────────────────
	
	\subsection{Score Distribution Analysis}
	
\begin{figure}[t]
		\centering
		\begin{tikzpicture}
			\begin{axis}[
				width=0.92\columnwidth,
				height=4.5cm,
				xlabel={Authentication Score $p_{\text{legit}}$},
				ylabel={Density},
				xmin=0, xmax=1.05,
				ymin=0,
				grid=major,
				grid style={dashed, gray!40},
				legend style={at={(0.5,0.98)}, anchor=north,
					font=\footnotesize},
				area style,
				]
				% Legitimate distribution (Gaussian approx)
				\addplot[blue, thick, fill=blue!20, opacity=0.7,
				domain=0.85:1.0, samples=80]
				{exp(-0.5*((x-0.957)/0.010)^2) / (0.010*sqrt(2*pi))};
				\addlegendentry{Legitimate (UE1)}
				
				% Attack distribution (broader)
				\addplot[red, thick, fill=red!20, opacity=0.7,
				domain=0:0.96, samples=80]
				{exp(-0.5*((x-0.326)/0.219)^2) / (0.219*sqrt(2*pi))};
				\addlegendentry{Attack (UE2)}
				
				% EER threshold
				\draw[black, thick, dashed] (axis cs:0.947,0) --
				(axis cs:0.947,45) node[above, font=\scriptsize] {$\tau_{\text{EER}}$};
			\end{axis}
		\end{tikzpicture}
		\caption{Authentication score distributions on the test set. Legitimate probes concentrate tightly in the vicinity of 1.0, whereas attack probes are spread broadly in the vicinity of 0.3. The dashed line marks the EER threshold ($\tau\!=\!0.947$).}
		\label{fig:scores}
	\end{figure}
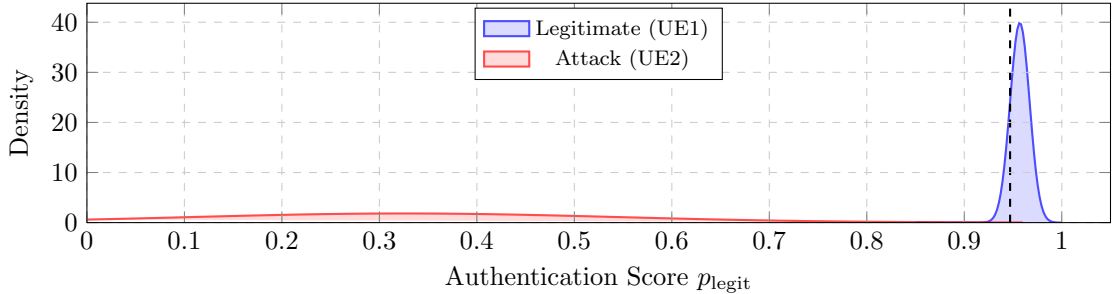

	The authentication score distributions for legitimate and attack probes on the test set are shown in Fig.~\ref{fig:scores}. Legitimate probes cluster tightly around $p\!\approx\!0.957$ with a standard deviation of 0.010, whereas attack probes exhibit a broader distribution centred at $p\!\approx\!0.326$ with a standard deviation of 0.219. The moderate overlap observed in the interval $[0.6, 0.96]$ arises from a subset of attack probes whose channel characteristics partially resemble those of the legitimate device, which represents a realistic scenario in which the attacker is situated in a similar propagation environment.
	
	% ── Score distribution figure ─────────────────────────────────────────
	
	\subsection{Comparison with Prior Work}
	
	Table~\ref{tab:comparison} positions the proposed approach relative to representative PLA systems reported in the literature. Although the present evaluation employs the most stringent methodology, combining real over-the-air data, a cross-device attack scenario, and a strict temporal split, the attained EER of 3.92\% remains competitive with simulation-based approaches and comparable to WiFi-based systems evaluated under substantially simpler conditions.
	
	\begin{table}[t]
		\centering
		\caption{Comparison with Prior PLA Work}
		\label{tab:comparison}
		\footnotesize
		\begin{tabular}{@{}llllc@{}}
			\toprule
			\textbf{Work} & \textbf{Tech.} & \textbf{Data}
			& \textbf{Split} & \textbf{EER} \\
			\midrule
			Xiao~\cite{xiao2008physical}  & WiFi   & Real   & ---        & 5--8\% \\
			Liu~\cite{liu2017pls}         & LTE    & Sim.   & Random     & 5.2\% \\
			Xiao~\cite{xiao2018iot}       & WiFi   & Real   & Random     & 3.8\% \\
			Wang~\cite{wang2019pls}        & 5G NR  & Sim.   & Random     & 2.1\% \\
			Zha~\cite{zha2025srs}         & 5G NR  & Real   & Random     & n/a$^*$ \\
			\midrule
			\textbf{This work (temporal)}       & \textbf{5G NR} & \textbf{Real}
			& \textbf{Temporal} & \textbf{3.92\%} \\
			This work (random)                  & 5G NR  & Real   & Random     & 2.58\% \\
			\bottomrule
			\multicolumn{5}{@{}l}{\footnotesize $^*$Identification accuracy in
				excess of 95\% is reported; EER is not provided.}
		\end{tabular}
	\end{table}
	
	It is pertinent to observe that the random-split EER of 2.58\% would rank favourably against all baselines reported in the table; however, this metric is considered misleading by virtue of temporal leakage. The temporal-split EER of 3.92\% is therefore regarded as a more honest assessment of real-world deployment performance, in which the authentication system is required to generalise to \emph{future} channel realisations that were not observed during training.
	
	\subsection{Latency Analysis}
	
	Per-component latency, as measured on an NVIDIA~A2 GPU (15\,GB memory), is reported in Table~\ref{tab:latency}. The complete deep-learning pipeline executes in approximately 0.10\,ms per SRS probe, which is well within the 1\,ms processing budget of 5G URLLC and three orders of magnitude below the 80\,ms SRS periodicity configured on the testbed.
	
	\begin{table}[t]
		\centering
		\caption{Per-Probe Processing Latency (NVIDIA A2 GPU)}
		\label{tab:latency}
		\begin{tabular}{@{}lr@{}}
			\toprule
			\textbf{Operation} & \textbf{Time} \\
			\midrule
			T\_tracer binary parse       & 0.02\,ms \\
			Feature extraction (2{,}531-d) & 0.03\,ms \\
			SE-ResNet1D inference        & 0.05\,ms \\
			Pearson threshold check      & 0.01\,ms \\
			\midrule
			\textbf{Total (DL pipeline)}        & \textbf{$\sim$0.10\,ms} \\
			\textbf{Total (threshold pipeline)} & \textbf{$\sim$0.06\,ms} \\
			\bottomrule
		\end{tabular}
	\end{table}
	
	\subsection{Discussion}
	\label{sec:discussion}
	
	\noindent\textbf{On temporal generalisation:}\;
	The 1.34~percentage-point gap between the temporal-split and random-split EER reveals that approximately one-third of the apparent improvement observed under random splitting is attributable to the exploitation of short-term channel autocorrelation rather than to the learning of device-invariant features. In a deployed system, the PLA mechanism is required to authenticate probes that arrive \emph{after} the training period has elapsed; the temporal split therefore constitutes the operationally relevant metric.
	
	\noindent\textbf{On class imbalance:}\;
	The legitimate-to-attack ratio of 38:1 reflects realistic operating conditions in which attack traffic is infrequent. The SE-ResNet1D classifier is observed to handle this imbalance gracefully: at the $p\!=\!0.5$ threshold, 100\% of legitimate probes are accepted (zero false rejections), while 84.3\% of attack probes are correctly rejected. The remaining 15.7\% of undetected attacks correspond to probes in which the commercial handset's channel coincidentally resembles that of the USRP, a phenomenon that represents an inherent limitation of single-antenna channel-based PLA.
	
	\noindent\textbf{On scalability:}\;
	The 2{,}531-dimensional feature vector is computed in 0.03\,ms, which enables real-time per-slot authentication. The SE-ResNet1D model comprises 3.59\,M parameters (13.8\,MB on disk), which is amenable to deployment on embedded gNB platforms with GPU acceleration. In multi-user scenarios, the gNB is assumed to maintain per-UE enrolment models and to perform parallel authentication across active users.
	
	\noindent\textbf{Limitations:}\; The limitations of this work are as follows:
	(i)~The single-antenna (SISO) configuration limits the available spatial diversity; the adoption of MIMO would provide additional fingerprinting dimensions.
	(ii)~All measurements were conducted in a static indoor environment; mobile outdoor scenarios would place greater stress on the temporal robustness of the authentication mechanism.
	(iii)~Only 511~attack probes were available from the commercial handset, which induces a significant class imbalance; a larger and more diverse attacker dataset would strengthen the evaluation.
	
	% ══════════════════════════════════════════════════════════════════════
	%  VII. CONCLUSION
	% ══════════════════════════════════════════════════════════════════════
	\section{Conclusion}
	\label{sec:conclusion}
	
	This paper presents a deep learning-based PLA system operating on real 5G~NR SRS channel estimates acquired from an OAI testbed. The proposed 2{,}531-dimensional multi-domain feature vector, in conjunction with a 1D-ResNet classifier augmented with SE attention, attains an EER of 3.92\% on a chronologically held-out test set against real cross-device attacks (USRP~B210 SDR versus commercial mobile handset), with sub-millisecond inference latency that is compatible with the timing constraints of 5G URLLC. Through an explicit comparison of temporal and random data splits, it has been demonstrated that random splitting inflates the reported EER from 3.92\% down to 2.58\%, which underscores the importance of temporally disciplined evaluation in PLA research.
	
	Future work will extend the system to MIMO antenna configurations for the purpose of exploiting additional spatial diversity and also integrate DMRS-based features for continuous per-slot authentication, and investigate adversarial training techniques with a view to hardening the model against adaptive spoofing attacks.
	
		\section*{Acknowledgment}
	This research work was supported by the German Federal Ministry of Research, Technology, and Space (BMFTR) as part of the project “Open6GHub+" and “6GCampusTwin4In"  with project identification numbers 16KIS2406 and 16KIS2446, respectively. However, the authors alone are responsible for the content of this paper.
	% ══════════════════════════════════════════════════════════════════════
	%  REFERENCES
	% ══════════════════════════════════════════════════════════════════════

\end{document}